\documentstyle [12pt,a4p,epsfig,amsmath,multicol]{article}
\begin{document}
\vspace*{0.6cm}

\begin{center} 
{\normalsize\bf Inter-charge forces in relativistic classical electrodynamics: electromagnetic
  induction in different reference frames}
\end{center}
\vspace*{0.6cm}
\centerline{\footnotesize J.H.Field}
\baselineskip=13pt
\centerline{\footnotesize\it D\'{e}partement de Physique Nucl\'{e}aire et 
 Corpusculaire, Universit\'{e} de Gen\`{e}ve}
\baselineskip=12pt
\centerline{\footnotesize\it 24, quai Ernest-Ansermet CH-1211Gen\`{e}ve 4. }
\centerline{\footnotesize E-mail: john.field@cern.ch}
\baselineskip=13pt
 
\vspace*{0.9cm}
\abstract{The force due to electromagnetic induction on a test charge
   is calculated in different reference frames. The Faraday-Lenz Law and different formulae for the fields
   of a uniformly moving charge are used. The classical Heaviside formula for the electric field of a moving
   charge predicts that, for the particular spatial configuration
   considered, the inductive force vanishes in the frame in which the magnet is in motion
   and the test charge at rest. In contrast, consistent results, in different frames, are given by the recently
   derived formulae of relativistic classical electrodynamics.}

\vspace*{0.9cm}
\normalsize\baselineskip=15pt
\setcounter{footnote}{0}
\renewcommand{\thefootnote}{\alph{footnote}}

 \par In the introduction of his 1905 paper on special relativity~\cite{Ein1} Einstein discussed
  the phenomenon of electromagnetic induction, discovered by Faraday, viewed either from a frame in which 
   the magnet is motion, or from one in which it is at rest. In this paper a careful re-analysis
   of this problem is performed in terms of the force on a test charge of magnitude $q$ in the vicinity
   of a magnet. The force on the charge, due to electromagnetic induction, is calculated in both
   the inertial frame, S, in which the magnet is a rest and the test charge is in motion
   as well as the frame, S', in which the magnet is in motion and the test charge is at rest. 
   \par Three different methods are used to perform the calculation:
    \begin{itemize}
    \item[(i)] Application of the Faraday-Lenz Law.
    \item[(ii)] Application of the Lorentz Force Law, using standard formulae
               of Classical Electromagnetism~\cite{JackHF,LLHF} (CEM) for the electric  
               and magnetic fields of a  uniformly moving charge.
\item[(iii)]  The formulae of Relativistic Classical Electrodynamics (RCED)
             ~\cite{JHF1,JHF2}, a covariant formalism developed recently by the present
               author, are used to calculate directly inter-charge forces.
    \end{itemize}
   The corresponding formulae are:
    \par \underline{Faraday-Lenz Law}
     \begin{equation}
       \Vec{F} = q \vec{E},~~~-\frac{1}{c}\frac{d \phi}{d t}= \int \vec{E}\cdot\ d\vec{s}
      \end{equation}
     \par \underline{CEM Formulae}
     \begin{eqnarray}
    \vec{E}(CEM) & = & \frac{Q \vec{r}}{r^3\gamma_u^2(1-\beta_u^2 \sin^2\psi)^{\frac{3}{2}}} =
                   \frac{Q (\hat{\imath}\cos \psi + \hat{\jmath}\sin \psi)}
                   {r^2\gamma_u^2(1-\beta_u^2 \sin^2\psi)^{\frac{3}{2}}} \\
    \vec{B}(CEM) & = & \frac{Q \vec{\beta_u} \times \vec{r}}{r^3\gamma_u^2(1-\beta_u^2 \sin^2\psi)^{\frac{3}{2}}} =   
    \frac{Q \beta_u \hat{k}\sin \psi}{r^2\gamma_u^2(1-\beta_u^2 \sin^2 \psi)^{\frac{3}{2}}} 
    \end{eqnarray}

     \par \underline{RCED Formulae}\footnote{In RCEM the forces between charges are calculated
            directly without the introduction of any field concept~\cite{JHF1}. For comparision
           purposes the terms in the force formula corresponding to the usual definitions of electric 
          and magnetic forces in Eqn(6) are expressed here in terms of corresponding electric and magnetic 
          fields $ \vec{E}(RCED)$ and $\vec{B}(RCED)$.} 
     \begin{eqnarray}
    \vec{E}(RCED) & = & \frac{Q \gamma_u}{r^3}[\vec{r} -\vec{\beta_u}(\vec{r} \cdot\vec{\beta_u})]
                       = \frac{Q}{r^2}\left(\frac{\hat{\imath} \cos \psi}{\gamma_u}+\gamma_u
            \hat{\jmath} \sin \psi \right) \\
    \vec{B}(RCED) & = & \frac{Q \gamma_u \vec{\beta_u} \times \vec{r}}{r^3} =   
    \frac{Q  \gamma_u \beta_u \hat{k}\sin \psi}{r^2} 
    \end{eqnarray}
     For the CEM and RCED calculations the force on the test charge is given by the Lorentz Force 
     Law:
    \begin{equation}
     \vec{F} = q (\vec{E}+\vec{\beta} \times \vec{B})
      \end{equation}
      where $\beta \equiv v/c$, $v$ is the speed of the test charge, and $c$ is the speed of light
       in vacuum. In Eqns(2-5) the 
    `source' charge of magnitude Q moves with uniform velocity $u \equiv \beta_u c$ along the x-axis, 
      $\gamma_u \equiv1/\sqrt{1-\beta_u^2}$, $\cos \psi = (\vec{v} \cdot \vec{r})/|\vec{v}||\vec{r}|$,
     $\vec{r}$ is the spatial vector connecting the source and test charges, and $\hat{\imath}$,
     $\hat{\jmath}$ and $\hat{k}$ are unit vectors parallel to the x-, y- and z-axes.
     \par In order to reduce the problem to its essentials, the `magnet' is constituted of just two equal 
      charges of magnitude Q with equal and opposite velocities $\vec{u}_+$ , $\vec{u}_-$ ,$|\vec{u}_+|
      = |\vec{u}_-| = u$ in the configuration shown in Fig.1a. The charges move parallel to the z-axis 
       and are situated at ($x$,$y$,$z$) = (0,$y$,0) and (0,-$y$,0), while the test charge is near to the
      symmetry point ($x$,0,0) and moves with velocity $\hat{\imath}v$ in the rest frame of the `magnet'
      constituted by the two source charges. Adding further moving charges, equidistant from the test charge,
      uniformly on a ring of radius $y$, to give a `one turn solenoid' complicates the evaluation
     of the fields and forces, but adds nothing to the essential dynamics of the problem.
        Since magnets are usually
     electrically neutral, the correspondence with a magnet constituted by an electron circulating
     in an atom or a one-turn solenoid would be made more exact by placing charges -Q, at rest in S,
      adjacent to the moving charges. Since however such charges produce no magnetic field in S,
      and an electric field at the test charge confined to the x-y plane in both S and S', the 
     following calculations of electromagnetic induction, where both electric and magnetic forces
     are parallel to the z-axis, is unchanged by the presence of such `neutralising' charges. They
      are therefore not considered in the following.
      \par In order to apply the Faraday-Lenz Law an imaginary rectangular current loop ABCD
        is drawn through the test charge in a plane perpendicular to the x-axis as shown in Fig.1a.
        If $a = {\rm AB} \ll b = {\rm BC}$, then, because of the symmetrical position of the
         loop, magnetic flux will, to a good approximation, cross only the short sides AB and DC as the loop
         attached to the test charge  
         moves through the field. In consequence, the line
       integral in (1) reduces to $ 2 E_z a$. Since (see Fig.1a) $\psi = \pi/2$, (3) or (5)
       give $\vec{B}(CEM) = \vec{B}(RCED)$ and the magnetic flux, $\phi$ threading the loop 
       ABCD is:
       \begin{equation}
        \phi(CEM)  = \phi(RCED) = ab[(B_+)_x+(B_-)_x] = \frac{2 abQ \gamma_u \beta_u \cos \theta}{r^2}
                                   = \frac{2 abQ \gamma_u \beta_u y}{r^3} 
    \end{equation}
     where $\vec{B}_+$ and $\vec{B}_-$ are the magnetic fields due to the charges with velocity
      $\vec{u}_+$ and $\vec{u}_-$ respectively. 
       Differentiating (7) w.r.t. $x$, noting that $v =dx/dt$, and using (1) gives:
       \begin{equation}
  -\frac{1}{c}\frac{d \phi}{d t}= 2 a E_z = \frac{6 abQ \gamma_u \beta_u \beta x y}{r^5}   
    \end{equation}
  So that the force on the test charge is:
      \begin{equation}
 F_z(FL) = q E_z = \frac{3 bqQ \gamma_u \beta_u \beta \cos \theta \sin \theta}{r^3}   
    \end{equation}
     where $\beta \equiv v/c$.
   \par The force on the test charge in S is now calculated using the Lorentz Force Law (6).
       Since both the CEM and RCED electric fields at the test charge lie in the x-y plane,
       only the magnetic force contributes in the z-direction. This force is given by the 
       y-component of $\vec{B}_++\vec{B}_-$ at the point ($x$,$b/2$,0). From the geometry of
       the x-y plane, shown in Fig.1b, and (3) or (5) with $\psi = \pi/2$:
     \begin{equation}
      B_y(x,\frac{b}{2},0) = (B_+)_y+(B_-)_y = Q \gamma_u \beta_u x \left(\frac{1}{r_+^3}
      -\frac{1}{r_-^3}\right)
     \end{equation}
       Assuming  then that $b \ll x,y$ it is found that:
     \begin{equation}
   \frac{1}{r_+^3}- \frac{1}{r_-^3} = \frac{3 b \cos \theta}{r^4} + O((b^2/r^5))
     \end{equation}
   so that from (10) and (11):
      \begin{equation}
           B_y(CEM) =  B_y(RCED) = \frac{3 b Q \gamma_u \beta_u \cos \theta \sin \theta}{r^3}
        + O((b^2/r^4))          
       \end{equation}
    Hence, using (6):
       \begin{equation}
           F_z(CEM) = F_z(RCED) =  \frac{3 b q Q \gamma_u \beta_u \beta \cos \theta \sin \theta}{r^3}
        + O((b^2/r^4)) 
       \end{equation}
      in agreement, to first order in $b$, with the Faraday-Lenz Law result (9).
\begin{figure}[htbp]
\begin{center}
\hspace*{-0.5cm}\mbox{
\epsfysize15.0cm\epsffile{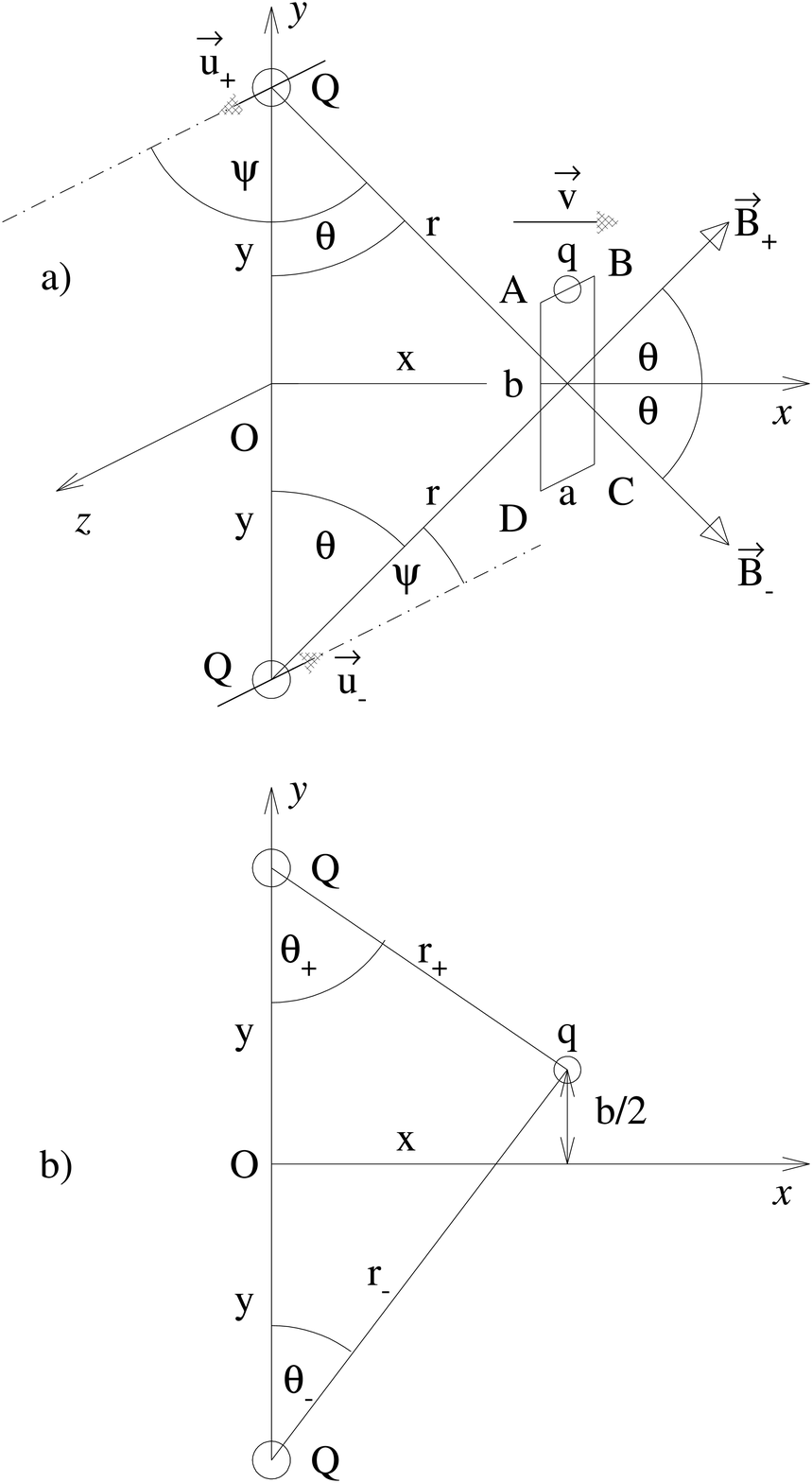}}   
\caption{{\sl Geometry for calculation of electromagnetic induction in the frame S in which the magnet
   is at rest and the test charge, of magnitude $q$, moves with velocity $\vec{v}$ parallel to the +ve x-axis.
    The `magnet' consists of two charges of magnitude $Q$ moving along the z-axis in opposite directions,
   each with speed $u$. The imaginary flux-loop ABCD is attached to the test charge. Various distances and angles
   are defined. $\vec{B}_+$ and  $\vec{B}_-$ are the magnetic fields at ($x$,0,0) generated by the
    charges of velocity  $\vec{u}_+$ and  $\vec{u}_-$. a) shows a perspective view and b) the x-y
    projection.}}
\label{fig-fig1}
\end{center}
 \end{figure}

\begin{figure}[htbp]
\begin{center}
\hspace*{-0.5cm}\mbox{
\epsfysize15.0cm\epsffile{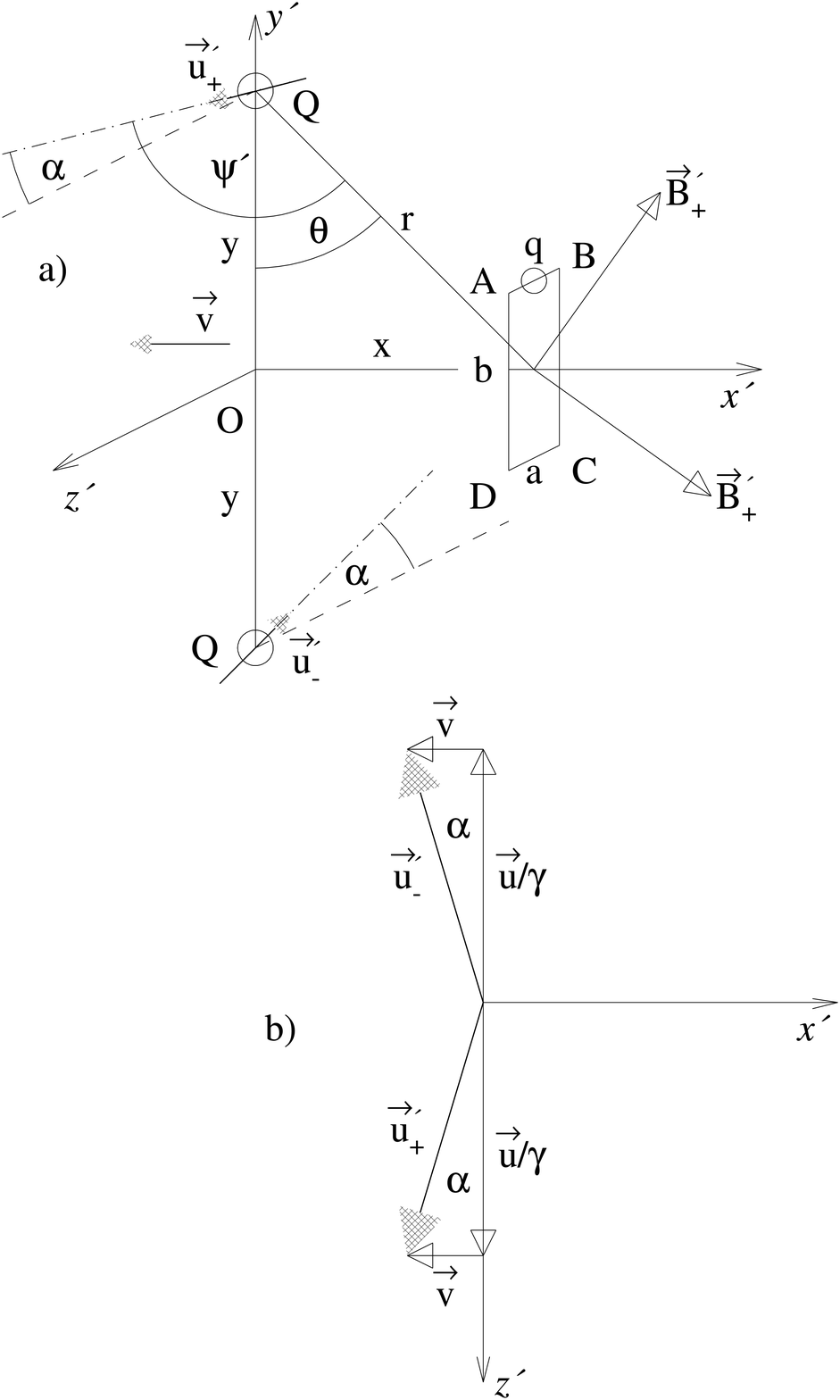}}   
\caption{{\sl Geometry for calculation of electromagnetic induction in the frame S' in which
 the test charge is at rest and the magnet moves at velocity $\vec{v}$ parallel to the -ve x-axis. Distances,
 angles, velocity vectors and magnetic fields are defined in a manner similar to those in Fig.1.
  a) shows a perspective view and b) the x'-z' projection.}}
\label{fig-fig2}
\end{center}
 \end{figure}

  \par The above calculations are now carried out in the frame S' where the magnet is in motion 
      and the test charge is at rest. Using Eqns(3) and (5) it follows from the geometry of Fig.2 that,
       the magnetic fluxes threading the loop ABCD in the frame S' are:
    \begin{eqnarray}
     \phi'(CEM) & =  & \frac{2 a b Q \gamma_{u'} \beta_u y}{\gamma r^3(1-\beta_{u'}^2 \sin^2 \psi' 
   )^{\frac{3}{2}}} = \frac{\gamma_{u'}}{\gamma \gamma_{u}(1-\beta_{u'}^2+\beta^2
      \sin^2 \theta)^{\frac{3}{2}}}\phi(CEM) \\
    \phi'(RCED) & =  & \frac{2 a b Q \gamma_{u'} \beta_u y}{\gamma r^3} =\frac{\gamma_{u'}}
    {\gamma \gamma_{u}}\phi(RCED)    
     \end{eqnarray}
    where, from the geometry of Fig.2b:
      \[ \beta_{u'} = \frac{\sqrt{\beta_u^2+ \gamma^2 \beta^2}}{\gamma},~~~\gamma_{u'}
        = \frac{1}{\sqrt{1- \beta_{u'}^2}},~~~\gamma = \frac{1}{\sqrt{1- \beta^2}}  \] 
     Note that each component of the spatial separation of the test and source charges remains 
     invariant under Lorentz transformation~\cite{JHF1,JHF3} so that, as shown in Fig.2,
      there is no distinction between the spatial interval $x$ and $x'$, $y$ and $y'$ and  $r$ and $r'$,
      or between the angles $\theta$ and $\theta'$.
      Since the calculation of the rate of change of the flux threading the loop is the same 
      whether the loop is displaced with velocity $v$ along the +ve x-axis as in S (Fig.1) or
       the source of the magnetic field is displaced with velocity $v$ along the -ve x-axis as in S' (Fig.2)
        the calculation of the z-component of the electric field using the Faraday-Lenz Law
        proceeds as above, with the results:
    \begin{eqnarray}
  F_{z'}(CEM) & =  & \frac{3 b qQ \gamma_{u'} \beta_u \beta \cos \theta  \sin \theta }
  {\gamma r^3(1-\beta_{u'}^2 \sin^2 \psi')^{\frac{3}{2}}} =
   \frac{\gamma_{u'}}{\gamma \gamma_{u}(1-\beta_{u'}^2+\beta^2 \sin^2 \theta)^{\frac{3}{2}}} F_z(CEM) \\
   F_{z'}(RCED) & =  & \frac{3 b qQ \gamma_{u'} \beta_u \beta \cos \theta \sin \theta}{\gamma r^3} =
  \frac{\gamma_{u'}}{\gamma \gamma_{u}} F_{z}(RCED)    
   \end{eqnarray}
       When the force calculations are performed in the frame S', by use of the Faraday-Lenz Law, consistent
     results are therefore, in general, no longer obtained. Only for the particular choice
      of the angle $\theta$ such that $\sin \theta = \beta_{u'}/\beta$ are the CEM and RCED predictions equal.
     \par Since the vectors $\vec{r}_+$,  $\vec{r}_-$ lie in the x'-y' plane, and the electric field in the CEM
      formula (2) is radial, it follows that the electric field at the test charge in S' also lies in this plane.
       Thus Eqn(2) predicts {\it no force, parallel to the z' axis, acts on the test charge in S'}. That is, that
       there is no effect of electromagnetic induction in this frame, in contradiction both the requirements
       of special relativity and the prediction of the Faraday-Lenz Law using either CEM or RCED fields.
      \par Finally the calculation is performed in the frame S' using the RCED electric field (4). 
          From the geometries of Fig.1b and Fig.2b:
     \begin{eqnarray}
      \vec{r}_+ & = & r_+(\hat{\imath} \sin \theta_+-\hat{\jmath} \cos \theta_+) \\                  
   \vec{r}_- & = & r_-(\hat{\imath} \sin \theta_++\hat{\jmath} \cos \theta_-) \\
  \vec{\beta}_+' & = & \beta_{u'}(-\hat{\imath} \sin \alpha+\hat{k} \cos \alpha) \\
 \vec{\beta}_-' & = & \beta_{u'}(-\hat{\imath} \sin \alpha-\hat{k} \cos \alpha) 
   \end{eqnarray}
    so that:
      \begin{eqnarray}
   \vec{r}_+\cdot  \vec{\beta}_+' & = & - \beta_{u'} r_+ \sin \theta_+ \sin \alpha
     = -  \beta_{u'} x \sin \alpha \\
  \vec{r}_-\cdot  \vec{\beta}_-' & = & - \beta_{u'} r_- \sin \theta_+ \sin \alpha
     = -  \beta_{u'} x \sin \alpha
   \end{eqnarray}
   Eqns(18), (19), (22), (23) and (4)  then give:
      \begin{eqnarray}
   E_{z'}(RCED) & = & (E_+)_z+ (E_-)_z = Q \gamma_{u'} \beta_{u'}^2  x  \sin \alpha \cos \alpha
        \left(\frac{1}{r_+^3}- \frac{1}{r_-^3}\right) \nonumber \\
     & = & \frac{3 b Q \gamma_{u'} \beta_{u'}^2 \cos \theta \sin \theta  \sin \alpha \cos \alpha}{r^3}
          +  O((b^2/r^4))
       \end{eqnarray}
    where Eqn(11) has been used.
  Hence:
     \begin{eqnarray}
    F_{z'}(RCED) &  = & q  E_{z'}(RCED)= \frac{3 b q Q \gamma_{u'} \beta_{u'}^2 \cos \theta
    \sin \theta  \sin \alpha \cos \alpha}{r^3} +  O((b^2/r^4)) \nonumber \\ 
 & = & \frac{3 b q Q \gamma_{u'} \beta_{u} \beta \cos \theta \sin \theta}{\gamma r^3}
          +  O((b^2/r^4)) \nonumber \\ 
 & = &  \frac{\gamma_{u'}}{\gamma \gamma_{u}} F_{z}(RCED)+  O((b^2/r^4))
      \end{eqnarray}
    where the relations following from the geometry of Fig.2b:
   \[ \sin \alpha = \frac{\beta}{\beta_{u'}},~~~\cos \alpha = \frac{\beta_u}{\gamma \beta_{u'}} \]
    have been used. 
    This result agrees, at first order in $b$, with that, (17), obtained by use of the Faraday-Lenz Law.
    \par The factor relating the inductive forces on the test charge in S and S' is:
    \begin{equation}
    \frac{\gamma_{u'}}{\gamma \gamma_{u}} = 1-\frac{1}{2}[\beta^2(\beta^2-\beta_u^2)+\beta_u^4]
               +O(\beta^2\beta_u^4,\beta_u^2 \beta^4)
   \end{equation}
     For $\beta \gg \beta_u$:
   \begin{equation}
    \frac{\gamma_{u'}}{\gamma \gamma_{u}} = 1+\beta^4 + O(\beta^6)
   \end{equation}
    while for   $\beta_u \gg \beta$:
  \begin{equation}
    \frac{\gamma_{u'}}{\gamma \gamma_{u}} = 1-\beta_{u}^4 + O(\beta_{u}^6)
   \end{equation} 
    so in these cases the forces in S and S' differ only by corrections of order the fourth power
    in the ratio of charge velocities to the speed of light.
     In summary, in the frame S, where the magnet is at rest, so that the magnetic fields are `static',
     and the test charge is in motion, all three methods of calculation yield the same result (9) or 
     (13) to the considered calculational accuracy. However when the Faraday-Lenz Law is used to 
      perform the calculation in the frame S' where the magnet is in motion and the test charge
     is at rest, the CEM result (16) is found to differ from the RCED one (17) by terms of O($\beta^2$).
     The CEM electric field formula (2) predicts the complete absence of electromagnetic induction
     in the frame S', in contradiction with the Faraday-Lenz Law predictions in this frame, and 
      with special relativity. The incompatibility of this formula, first derived by Heaviside~\cite{Heaviside},
      more than a decade before the advent of special relativity, with the requirements 
     of the latter has been previously demonstrated by comparing calculations of Rutherford scattering
     in different inertial frames~\cite{JHF2} as well as by Jackson's `Torque Paradox'~\cite{JackTP},
      which is resolved~\cite{JHF2} by the use of the RCED force formula that is the combination of (4),(5) and (6).
     \par All four results (9),(13),(17) and (26) of the calculations of the force, using the RCED formulae, give
      consistent results. The forces in the frames S and S' are found to differ only by corrections
     of order the fourth power in the ratio of velocities to the speed of light. The forces are different
     due to the relativistic time dilatation effect which results in different accelerations in 
     different inertial frames. That forces are different in different inertial frames is already evident
     from inspection of Eqns(4) and (5) by comparing the fields in the frame where the source charge is at rest
     ($\beta_u = 0$, $\gamma_u = 1$) with those shown for an arbitary value of $\beta_u$.

     \par Since the original version of this paper was written some two years ago, convincing experimental
      evidence has been obtained~\cite{Ketal1} for the non-retarded nature of electrodynamical
      force fields, as in RCED. The relation of these results to the basis of RCED in QED has also been
      discussed~\cite{JHFnrf}.



\begin{thebibliography}{99}
\bibitem{Ein1} 
A.Einstein, Annalen der Physik {\bf17}, 891 (1905). \newline
  English translation by W.Perrett and G.B.Jeffery in `The Principle of Relativity'
  (Dover, New York, 1952) P37, or in `Einstein's Miraculous Year'
  (Princeton University Press, Princeton, New Jersey, 1998) P123.
\bibitem{JackHF}
J.D.Jackson, `Classical Electrodynamics', 3rd Ed (Wiley, New York, 1998) P558,560.
\bibitem{LLHF}
L.D.Landau and E.M.Lifshitz, `The Classical Theory of Fields',
   2nd Edition (Pergamon Press, Oxford, 1962) Section 38, P103.
\bibitem{JHF1}
J.H.Field, Phys. Scr. {\bf 74} 702 (2006),
 http://xxx.lanl.gov/abs/physics/0501130.
\bibitem{JHF2}
J.H.Field, Int. J. Mod. Phys. A Vol 23 No 2 327 (2008),
http://xxx.lanl.gov/abs/physics/0507150. 
\bibitem{JHF3}
J.H.Field, Phys. Scr. {\bf 73} 639 (2006),
 http://xxx.lanl.gov/abs/physics/0409103.
\bibitem{Heaviside}
O.Heaviside, The Electrician,  {\bf 22} 1477 (1888),
Philos. Mag. {\bf 27} 2324 (1889).
\bibitem{JackTP}
 J.D.Jackson, Am. J. Phys. {\bf 72} 1484 (2004).
\bibitem{Ketal1}
A.L.Kholmetskii {\it et al.} J. Appl. Phys. {\bf 101} 023532 (2007).
\bibitem{JHFnrf}
 J.H.Field, `Quantum electrodynamics and experiment demonstrate
   the non-retarded nature of electrodynamical force fields',
   http://xxx.lanl.gov/abs/0706.1661.
\end{thebibliography}
\end{document}